\documentclass[sigconf, nonacm]{acmart}

\newcommand\vldbdoi{XX.XX/XXX.XX}
\newcommand\vldbpages{XXX-XXX}
\newcommand\vldbvolume{17}
\newcommand\vldbissue{1}
\newcommand\vldbyear{2024}
\newcommand\vldbauthors{\authors}
\newcommand\vldbtitle{\shorttitle} 
\newcommand\vldbavailabilityurl{URL_TO_YOUR_ARTIFACTS}
\newcommand\vldbpagestyle{plain} 

\usepackage{pifont}
\usepackage[ruled,vlined,resetcount]{algorithm2e}
\usepackage{makecell}

\newcommand{\ourwork}{\textsc{D2PC}\xspace}

\newcommand{\gddb}{geo-distributed\xspace}
\newcommand{\GDDB}{Geo-Distributed\xspace}
\newcommand{\layered}{layered mode\xspace}
\newcommand{\Layered}{Layered mode\xspace}
\newcommand{\LAYERED}{Layered Mode\xspace}
\newcommand{\integrate}{co-designed mode\xspace}
\newcommand{\Integrate}{Co-designed mode\xspace}

\newcommand{\prepare}{\texttt{Prepare}\xspace}
\newcommand{\commit}{\texttt{Commit}\xspace}
\newcommand{\abort}{\texttt{Abort}\xspace}
\newcommand{\preparestate}{\texttt{Prepared}\xspace}
\newcommand{\precommit}{\texttt{PreCommit}\xspace}
\newcommand{\commitstate}{\texttt{Commit}\xspace}
\newcommand{\executed}{\texttt{Executed}\xspace}
\newcommand{\abortstate}{\texttt{Abort}\xspace}
\newcommand{\preparereq}{\texttt{Prepare}\xspace}

\newcommand{\majority}{$ \mathcal{F} $\xspace}
\newcommand{\cross}{inter-datacenter\xspace}

\newcommand{\intra}{intra-datacenter\xspace}

\newcommand{\cc}{concurrency control period\xspace}
\newcommand{\CC}{Concurrency Control Period\xspace}
\newcommand{\Cc}{Concurrency control period\xspace}

\newcommand{\primary}{correspondent coordinator\xspace}

\newcommand{\PRIMARY}{Correspondent Coordinator\xspace}
\newcommand{\co}{co-coordinator\xspace}
\newcommand{\Co}{Co-coordinator\xspace}
\newcommand{\coss}{co-coordinators\xspace}
\newcommand{\Cos}{Co-coordinators\xspace}
\newcommand{\lock}{2PC+2PL\xspace}
\newcommand{\occ}{2PC+OCC\xspace}
\newcommand{\dlock}{D2PC+2PL\xspace}
\newcommand{\docc}{D2PC+OCC\xspace}

\newcommand{\locknro}{2PC+2PL-NRO\xspace}
\newcommand{\ourworknro}{D2PC+2PL-NRO\xspace}

\newcommand{\war}{\emph{rw}\xspace}
\newcommand{\raw}{\emph{wr}\xspace}
\newcommand{\waw}{\emph{ww}\xspace}

\newcommand{\zzh}[1]{{\color{black}{#1}}}

\begin{document}
\title{Fast Commitment for Geo-Distributed Transactions via Decentralized \Cos}
\author{Zihao Zhang}
\affiliation{%
  \institution{East China Normal University}
}
\email{zihaozhang@stu.ecnu.edu.cn}

\author{Huiqi Hu} \authornote{represents the corresponding author.}
\affiliation{%
  \institution{East China Normal University}
}
\email{hqhu@dase.ecnu.edu.cn}


\author{Xuan Zhou}
\affiliation{%
  \institution{East China Normal University}
}
\email{xzhou@dase.ecnu.edu.cn}

\author{Yaofeng Tu}
\affiliation{
    \institution{ZTE Corporation}
}
\email{tu.yaofeng@zte.com.cn}

\author{Weining Qian}
\affiliation{%
  \institution{East China Normal University}
}
\email{wnqian@dase.ecnu.edu.cn}

\author{Aoying Zhou}
\affiliation{
    \institution{East China Normal University}
}
\email{ayzhou@dase.ecnu.edu.cn}

\begin{abstract}
In a geo-distributed database, data shards and their respective replicas are deployed in distinct datacenters across multiple regions, enabling regional-level disaster recovery and the ability to serve global users locally. However, transaction processing in geo-distributed databases requires multiple cross-region communications, especially during the commit phase, which can significantly impact system performance. 


To optimize the performance of geo-distributed transactions, we propose Decentralized Two-phase Commit (\ourwork), a new transaction commit protocol aiming to minimize the negative impact of cross-region communication. In \ourwork, we employ multiple co-coordinators that perform commit coordination in parallel. Each \co is responsible for collecting 2PC votes and making a \precommit decision in its local region. This approach allows for the concurrent invocation of multiple cross-region network round trips, and each region can conclude its concurrency control locally before replication is complete, thus significantly reducing the chances of blocking and enhancing system concurrency. Moreover, we propose the bypass leader replication reply method, leveraging decentralized \coss to bypass the leader for message transmission, thereby reducing the commit latency. Experimental results have demonstrated that \ourwork can reduce commit latency by 43\% and improve throughput by up to 2.43 $\times$ compared to the existing alternative geo-distributed transaction processing methods.
\end{abstract}

\maketitle
\vspace{-2mm}
\pagestyle{\vldbpagestyle}
\begingroup\small\noindent\raggedright\textbf{PVLDB Reference Format:}\\
\vldbauthors. \vldbtitle. PVLDB, \vldbvolume(\vldbissue): \vldbpages, \vldbyear.\\
\endgroup
\begingroup
\renewcommand\thefootnote{}\footnote{\noindent
This work is licensed under the Creative Commons BY-NC-ND 4.0 International License. Visit \url{https://creativecommons.org/licenses/by-nc-nd/4.0/} to view a copy of this license. For any use beyond those covered by this license, obtain permission by emailing \href{mailto:info@vldb.org}{info@vldb.org}. Copyright is held by the owner/author(s). Publication rights licensed to the VLDB Endowment. \\
\raggedright Proceedings of the VLDB Endowment, Vol. \vldbvolume, No. \vldbissue\ %
ISSN 2150-8097. \\
\href{https://doi.org/\vldbdoi}{doi:\vldbdoi} \\
}\addtocounter{footnote}{-1}\endgroup

\vspace{-4mm}
\ifdefempty{\vldbavailabilityurl}{}{
\vspace{.3cm}
\begingroup\small\noindent\raggedright\textbf{PVLDB Artifact Availability:}\\
The source code, data, and/or other artifacts have been made available at \url{\vldbavailabilityurl}.
\endgroup
}
\section{Introduction}\label{introduction}

\GDDB databases have become a vital infrastructure for hosting cross-region applications, such as international banking, popular e-commerce platforms, social media, etc. Prominent examples of \gddb databases include Spanner~\cite{spanner1} and CockroachDB~\cite{cockroachdb}. These databases employ the strategy of partitioning data into shards and replicating them across datacenters globally.

The competitiveness of a \gddb database relies on its ability to handle \gddb transactions that span multiple regions. To ensure the atomicity of distributed transactions, it is necessary to employ an atomic commitment protocol like Two-phase Commit (2PC) to coordinate the commit phases. Additionally, to ensure high availability, a consensus protocol should be used to maintain consistency among cross-region data replicas. However, both commitment and consensus protocols involve multiple rounds of cross-region communication, each of which can introduce significant latency, lasting hundreds of microseconds. This extended latency can prevent critical applications from meeting their required service level agreements (SLA). It can also increase lock holding time and thus the chances of contention, posing a serious threat to overall performance.
Therefore, to make \gddb transactions practical, it is essential to minimize the negative impact of cross-region communication on transaction processing.

\zzh{
In most existing \gddb database systems, the transaction layer, which runs the commitment protocol, is typically built upon the replication layer that hosts the consensus protocol. A typical example is Spanner~\cite{spanner1, spanner2}, which employs the 2PC protocol layered over the Multi-Paxos~\cite{multi-paxos} protocol. 
However, in the \gddb deployment where shard leaders span different regions, distributed transaction committing has to bear the cost of multiple rounds of cross-region communication. This often forces upper-layer applications to give up using \gddb transactions for more sustainable performance.}



Several recent research works~\cite{tapir, janus, carousel, mdcc, ov, starry} have proposed tightly integrating consensus protocols and commit protocols to minimize the number of network round trips needed for transaction commitment. This integration aims to reduce latency by allowing all replicas to process transactions concurrently, in contrast to the layered architecture that relies on the leader for transaction processing.
In non-conflict scenarios, these approaches typically require only a single network round trip to commit a transaction, resulting in significantly reduced latency. However, in the presence of conflicts, the impact of cross-region communication on performance remains outstanding. On one hand, these approaches still require a minimum of two network round trips to commit a conflicting transaction. On the other hand, multiple cross-region communications may still occur long \cc, leading to high contention or abort rates.
Meanwhile, involving all replicas in transaction processing is fundamentally deviating from the architecture in existing databases that require a leader node for transaction processing. 

\zzh{
In this work, we introduce Decentralized Two-phase Commit (\ourwork), a novel commit protocol tailored for the common layered architecture in \gddb databases.
The primary objective of \ourwork is to mitigate the impact of cross-region communication on concurrency, thereby enhancing the overall throughput of \gddb transactions. Additionally, \ourwork also works to minimize the effects of cross-region communication on transaction latency, addressing both non-conflict and conflict scenarios. }


In the layered architecture, we have observed that the processes of 2PC protocol and replication require multiple rounds of cross-region communication. To reduce the overhead, \ourwork decouples and parallelizes these processes. Furthermore, during the commit phase, we have identified that the single coordinator frequently incurs cross-region communication. Consequently, in \ourwork, we aim to distribute the commit coordination duty among regions, thereby each server can communicate with a coordinator locally.
 
In a nutshell, \ourwork deploys a set of \coss across all datacenters. 
Each \co operates independently and is responsible for collecting votes from participant shards and making a \precommit decision without waiting for replication completion. 

\zzh{Upon reaching the \precommit stage, the \co promptly notifies the local participant leaders to conclude concurrency control.  This action effectively reduces the concurrency control duration in the commit phase to 0.5 cross-region round trip. Besides, during the commit phase, each \co directly relays the replication reply from its co-located followers to the \primary, bypassing intermediate communication with the shard leader. This significantly reduces the commit latency to 1 - 1.5 cross-region round trips. Consequently, \ourwork outperforms existing alternative methods in both concurrency control duration and latency.}

The contributions of this paper can be summarized as follows:
\begin{itemize}
\vspace{-3mm}
\zzh{
    \item We observed that the single coordinator in the 2PC protocol can lead to additional cross-region communication during transaction commits. To eliminate it, we decentralized the coordinator into a group of \coss distributed across every datacenter.


    \item Building upon \coss, we proposed the decentralized transaction commitment which can minimize the impact of cross-region communication on concurrency and commit latency for geo-distributed transactions.}


    \item \zzh{We conducted extensive experiments in a multi-cloud scenario, using the benchmarks of Retwis and YCSB+T. We combined \ourwork with OCC and 2PL respectively. \docc improve the throughput by up to 2.43 $\times$ than \occ. Similarly, demonstrates an improved throughput of 1.73 $\times$ compared to \lock, and this improvement increases to 2.33 $\times$ when read optimization is enabled.}


    \vspace{-2mm}
\end{itemize}
   
The rest of the paper is organized as follows. $\S$ \ref{background} introduces the background and motivations of \ourwork. $\S$ \ref{design} sketches ideas and the architecture of \ourwork. $\S$~\ref{protocol} presents \ourwork in greater detail, elaborating on its commit protocol and the techniques to shorten \cc. \S~\ref{evaluation} reports the experimental results. Finally, we conclude the paper in \S~\ref{conclusion}.
\section{Background and Related Work}\label{background}

\subsection{Two-phase Commit and the Blocking Issue}\label{background1}

In distributed database management systems (DDBMS), data is partitioned into shards to achieve scalability. However, transactions that span multiple shards necessitate an atomic commitment protocol to ensure atomicity. This protocol ensures that all involved servers achieve a consensus on the commit decision.

Two-phase commit (2PC) is widely used in many DDBMSs to ensure atomicity. 2PC consists of two phases: the Prepare phase and the Commit phase. In the Prepare phase, participants send votes to the coordinator, who can proceed to the Commit phase only if all participants vote to commit. However, the 2PC protocol faces a blocking issue. As only one coordinator can make the final commit decision, if the coordinator fails before notifying the decision, participants may be blocked until the coordinator recovers. 
This block prevents the release of transaction resources, such as locks, impeding system progress.

Some solutions have been proposed to address the blocking issue.
3PC~\cite{3pc} introduces a new phase called the "prepared to commit" phase. In this phase, all participants must acknowledge the commit decision before actually committing. This ensures that all participants, not just the coordinator, are aware of the decision. \zzh{E3PC~\cite{e3pc} further enhances the availability of 3PC by introducing a quorum.} Paxos commit~\cite{paxoscommit} combines Paxos~\cite{paxos} with 2PC, using Paxos to replicate the commit decision to a set of replicas. Easy Commit~\cite{easycommit} mandates participants to forward the coordinator's decision to all others before committing. Cornus~\cite{cornus} is a one-phase commit protocol designed for disaggregated-storage architecture with the assumption that the storage is high-available. 

In summary, these protocols ensure fault-tolerant commit decisions to overcome the blocking issue. Following this principle, we deploy a set of \coss to tolerate failures, simplifying the solution of the blocking issue without compromising transaction processing performance.



\vspace{-2mm}
\subsection{Transaction Commit in \GDDB  Databases}\label{background2}

Nowadays, high availability is a crucial requirement for many applications, leading databases to adopt consensus protocols like Paxos~\cite{paxos} or Raft~\cite{raft} for fault-tolerant replication of shards across multiple replicas. Consequently, in DDBMSs, the transaction commits necessitate the combination of 2PC and consensus protocols to ensure correctness. The collaboration between these two protocols can be summarized in two patterns.

\noindent \textbf{\Layered.} 
Many commercial \gddb databases, including Spanner~\cite{spanner1, spanner2}, CockroachDB~\cite{cockroachdb}, and TiDB~\cite{tidb}, construct their transaction layer, encompassing the 2PC protocol and concurrency control protocol, atop leader-based consensus protocols. In layered architecture, the transaction layer necessitates 2PC coordination for commit decisions, while consensus protocols require coordination for maintaining consistent transaction order among replicas. However, this over-coordination introduces significant overhead, prompting efforts to optimize it. For instance, CockroachDB introduces parallel commit to achieve single-round commit latency, but this assumes that the coordinator and shard leaders are co-located, which doesn't align with geo-distributed shard scenarios. CockroachDB also proposes future-time transactions~\cite{cockroach22} for fast reading in \gddb deployments, but it requires bounded clock skew and compromises write latency. Both of them are not general solutions to address over-coordination in \gddb transactions.

\begin{figure}[t]
	\centerline{\includegraphics[width=7cm]{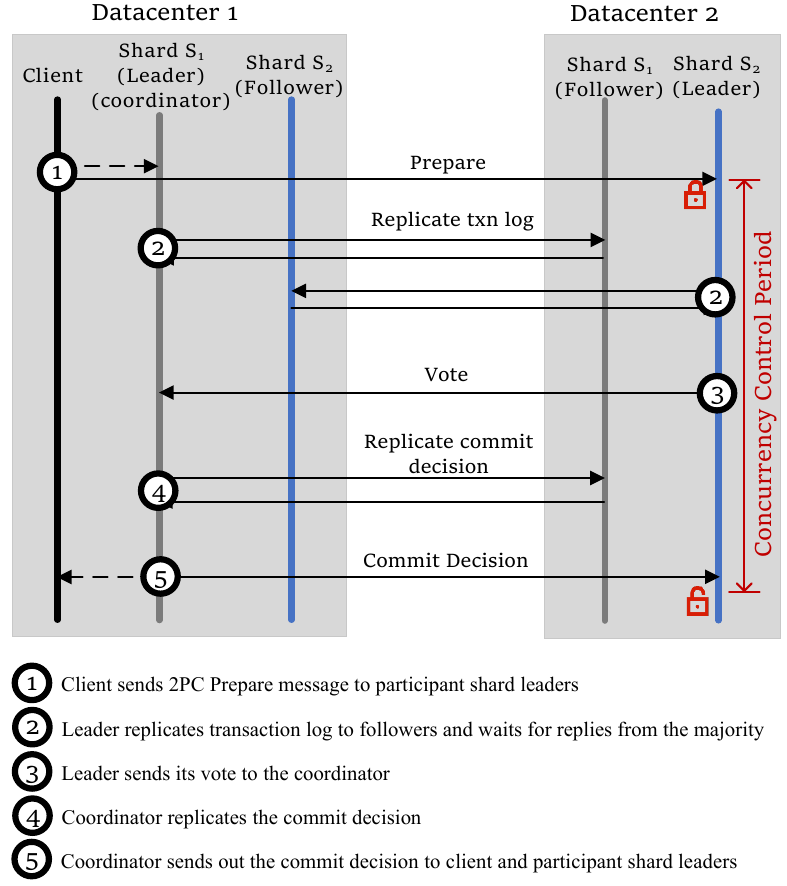}}
	\vspace{-2mm}
	\caption{\zzh{Example of committing a transaction in a Spanner-like protocol. Solid and dashed arrows represent inter-datacenter and intra-datacenter messages, respectively.}} \label{2pc}
	\vspace{-4mm}
	
\end{figure}

To be specific, we give an example in Fig.~\ref{2pc} to demonstrate the performance issues caused by over-coordination.
To simplify the explanation, we only consider two data shards, S$_{1}$ and S$_{2}$, each with two replicas located in different datacenters. In practice, there are typically many data shards, each with at least three replicas. To commit a transaction, a coordinator (the leader of S$_{1}$) is selected among the participant shard leaders. 
The process begins with the client sending a \prepare message to leaders (Process \ding{192}). The leader validates whether the transaction can be committed and replicates the transaction log to the follower replicas (Process \ding{193}). After receiving acknowledgments from a majority of followers, the leader sends its vote to the coordinator (Process \ding{194}). 
The coordinator collects votes from all participant shard leaders and decides whether to commit or abort. The commit decision is also replicated for fault tolerance (Process \ding{195}). Finally, the coordinator sends a \commit message to the client. Therefore, committing transactions incurs a total of 3 \cross round-trip times (RTTs).

Over-coordination also causes a long \cc. The \cc refers to the duration in which a transaction can impact others. Recall the example in Fig.~\ref{2pc}, the \cc starts when the participant leader receives the \prepare message (e.g., the leader acquires locks on operating records). Then the \cc concludes when the \commit message is received (e.g., the leader release locks). Consequently, the \cc extends over a duration of 3 \cross RTTs. This extended duration of exclusive resource access during the \cc severely restricts the concurrency of the database system. 

In practice, the votes of each shard are replicated to a majority of replicas, ensuring that in the event of coordinator failure, the commit decision can be recovered by obtaining votes from the replicas of each participant. 
This eliminates the need for replicating the commit decision (Process \ding{195} in Fig~\ref{2pc}). As a result, both the commit latency and \cc length can be reduced to 2 \cross RTTs.

\vspace{1mm}

\begin{figure*}[t]
	\centerline{\includegraphics[width=19cm]{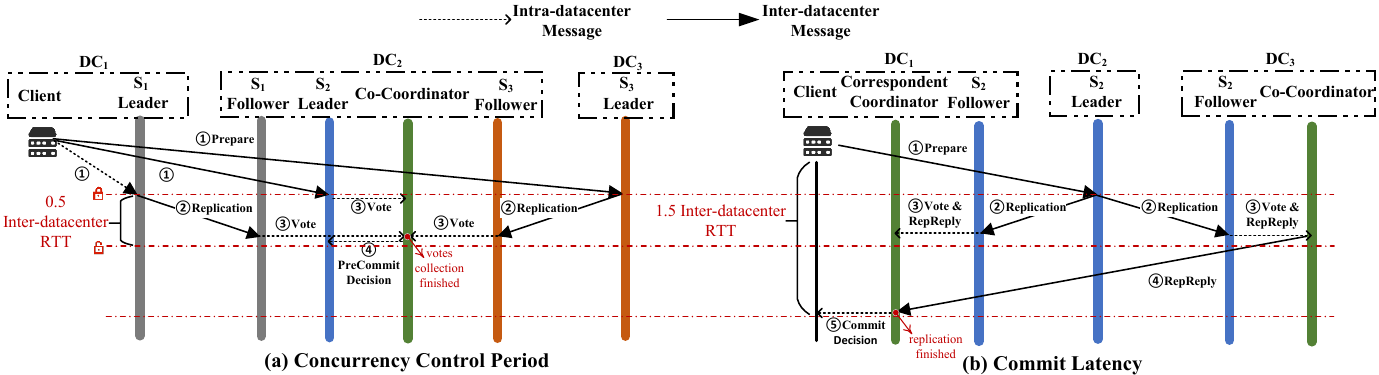}}
	\vspace{-4mm}
	\caption{Example of how to shorten the \cc and reduce the commit latency in \ourwork.}
	\vspace{-4mm}
	\label{overview}
\end{figure*}

\noindent \textbf{\Integrate.}
To solve the high commit latency issue, many works co-design 2PC and replication protocols. Replicated Commit~\cite{replicatedcommit} conversely builds Paxos over 2PC. This allows each datacenter to make a commit decision independently and then use Paxos to reach a consensus among datacenters. By delivering 2PC messages within the datacenter, round trips across datacenters are reduced. MDCC~\cite{mdcc}, TAPIR~\cite{tapir}, Janus~\cite{janus}, OV~\cite{ov}, Starry~\cite{starry}, and Carousel~\cite{carousel} integrate consensus and transaction management to achieve fast commitment. In these protocols, the \prepare message of a transaction is replicated to all replicas of every participant shard. Each replica then decides whether it can be committed. The coordinator collects all replies and makes the final decision. In the ideal conflict-free scenario, a transaction can be committed in one \cross RTT. But when conflict occurs, extra costs are incurred to resolve conflicts. For example, MDCC and TAPIR will abort conflicting transactions, while Janus, Starry, and Carousel require extra communication to reach a consistent order. \zzh{GeoGauss~\cite{geogauss} utilizes multi-master transaction processing to achieve a similar goal, it mainly focuses on the efficient coordination-free conflict resolution mechanism.}

Furthermore, deterministic databases~\cite{aria, calvin, bohm, qstore, ThomsonA10} handle 2PC and replication protocols in distinctive manners. 
These databases focus on ensuring consistency and persistence of transaction inputs. They require all replicas to deterministically execute the same input transactions to achieve consistent results. 

In summary, the optimization goal of \integrate is to reduce the commit latency, without specific emphasis on reducing the length of \cc.
The fundamental concept behind \integrate diverges from the traditional approach of relying on a single leader for transaction processing. Instead, it enables all replicas to execute transaction logic, necessitating significant customization of the transaction and storage layers. This customization poses a significant challenge for seamless integration into existing databases.

In this paper, our optimized approach focuses on addressing the challenges posed by the long \cc length and high commit latency issues separately.  Our objective is to achieve a significant reduction in \cc length and a comparable reduction in commit latency, similar to what has been achieved in \integrate. 
Our optimizations are specifically designed for the \layered architecture that relies on leader-based transaction processing, ensuring compatibility with existing database infrastructures.

\subsection{Timing of Write to be Visible}\label{background3}


Besides the high commit latency prolonging \cc, another reason for long \cc comes from \emph{delayed write visibility}, which is used to ensure \emph{recoverability} of transactions.
Consider a transaction $T_{1}$ write on data $x$, then transaction $T_{2}$ reads $T_{1}$'s write on $x$, the system must ensures that $T_{2}$ is committed after $T_{1}$, thus preventing $T_{1}$ from aborting after $T_{2}$ is committed. The schedule is \emph{recoverable}~\cite{recoverable} if the commit order does not violate the order of read-after-write dependencies. To ensure recoverability, most concurrency control mechanisms delay the visibility of writes until the transaction is committed, resulting in a long \cc.

Many works have investigated \emph{early write visibility} in concurrency control protocol, which enables transactions to read uncommitted writes, and then forces the commit order to be consistent with the read-after-write order to maintain recoverability~\cite{recoverable, swv1, swv2, swv3}. Most of them are designed for locking-based protocols. For example, ELR~\cite{elr1, elr2, elr-use1, elr-use2} allows the transactions that have finished execution to release locks before logging. CLV~\cite{clv} is proposed for the same goal and it re-designs the lock table to track and enforce the commit order of dependent transactions. Bamboo~\cite{bamboo} explores violating 2PL by allowing lock release during execution to further enhance concurrency. There are also some works that employ early write visibility in non-locking protocols. PWV~\cite{pwv} is designed for deterministic databases and leverages the determinism that orders transactions before execution. This allows updates made by transactions to be visible before their execution is completed. Hekaton~\cite{hekaton1, hakaton2} proposes a protocol for multi-version concurrency control that allows uncommitted dirty data to be read if it is in the "preparing" state.



However, research on applying early write visibility for distributed transactions, especially in the realm of \gddb transactions, is sparse. DLV~\cite{dlv} is the most relevant existing work in this domain, investigating various timings to perform lock violation. Given the substantial impact of early write visibility on mitigating the \cross communication penalty in \gddb transactions, \ourwork is dedicated to concluding concurrency control early. The opportunity for early write visibility in \ourwork arises from decentralized commit coordination. 
As detailed in \S~\ref{design_decentralized}, each datacenter independently makes \precommit decisions, eliminating the communication delay with the centralized coordinator inherent in traditional 2PC. \ourwork also separates commit coordination from replication, thereby excluding the message delay of replication from \cc.
As a result, \ourwork stands out as a comprehensive commit protocol capable of reducing both \cc length and commit latency, and it remains compatible with both 2PL and OCC, these differences distinguish it from prior work on early write visibility.

\section{Design Overview}\label{design}

In this section, we provide a concise overview of the design of \ourwork, highlighting the rationale behind decentralized commit, the parallelization of processes, and the strategies employed to minimize \cc length and reduce commit latency. Additionally, we will introduce the transition of transaction state and the architecture of \ourwork.


\vspace{-2mm}
\subsection{Decentralized Commit and Process Decoupling}\label{design_decentralized}

The location of the coordinator significantly impacts the length of \cc. Specifically, when the coordinator is located in a different region than the participant leader, the communication between them introduces \cross communication, resulting in message delays and extended waiting times for participant leaders to conclude the \cc.

We observe that this issue primarily arises from the reliance on a single coordinator in the 2PC protocol, which makes the location of the coordinator critical. To address this, we propose a shift from a centralized to a decentralized commit coordination pattern. This involves deploying a set of \emph{\coss} across all datacenters. By doing so, each server can be co-located with a \co, enabling \intra communication between the participant leader and the \co. As a result, the waiting time for the commit decision on participant leaders can be reduced.

Two essential steps need to be completed before the coordinator can decide to commit: 2PC vote collection (receiving votes from all participants), and replication (where all participant leaders replicate the transaction log). As illustrated in Fig.~\ref{2pc}, the two steps are interleaved, with processes \ding{192} and \ding{194} for 2PC, and process \ding{193} for replication. The interleaving of processes results in all message delays being included in the waiting time for participant leaders to receive the commit decision.

\zzh{However, we observe that the processes of 2PC and replication are independent of each other.  Hence, we decouple the 2PC and replication processes and illustrate them in Fig.~\ref{overview}. In this figure, (a) depicts the reduction in \cc length, and (b) shows the optimization of commit latency (details in \S~\ref{design_bypass leader}). }


\zzh{The core idea of \ourwork is that \emph{if the 2PC's vote collection process occurs in parallel with the replication process, then the vote collection can be completed before replication with the assistance of the \co.}
As shown in Fig.~\ref{overview}, the second dashed red line can be regarded as the completion of vote collection, and the third dashed red line represents the completion of replication. Therefore, we intend to \textit{decouple and parallelize the processes of 2PC and replication, allowing each datacenter to collect votes and make a \precommit decision early via the \co,} thereby shortening the length of \cc. }


\zzh{To be specific, in Fig.~\ref{overview}(a), when a participant leader receives the \prepare message(message \ding{172}), it replicates the transaction log and its vote to followers (message \ding{173}). Each follower then forwards the vote to the co-located \co (message \ding{174}), e.g., the followers of S$_{1}$ and S$_{3}$ send the vote to the co-located \co in DC$_{2}$ upon receiving the replication message from their respective leaders. Upon the \co collects votes from the co-located replicas, it can make \precommit decision before the replication is completed, and promptly notify the co-located participant leaders about the decision (message \ding{175}).}

\zzh{
At this stage, the global serial order of transactions has already been determined, as all participant shards have agreed to commit. Therefore, concluding \cc at this point not only excludes the bother of aborts caused by violating serializable, but also achieves a much shorter \cc length. Considering that, participant leaders conclude \cc early by \textit{enabling earlier writes visible. }}  

\zzh{As shown in Fig.~\ref{overview}(a), when leader receives the \prepare message (message \ding{172}), \cc starts. Then when the \co receives votes of each participant shards (message \ding{174}), the \cc can be concluded. Therefore, the \cc length is only 0.5 \cross RTT. }
 
\begin{figure}[t]
	\centerline{\includegraphics[width=8cm]{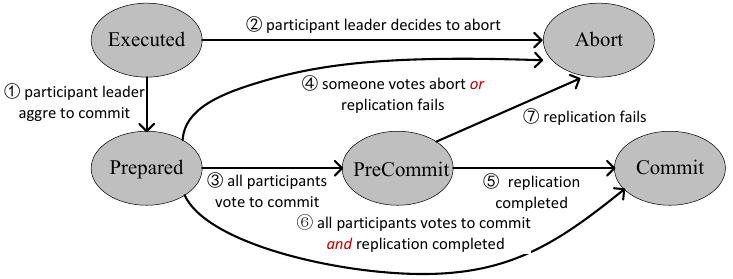}}
	\vspace{-2mm}
	\caption{\zzh{Transition Graph of Transaction State.}}
	\vspace{-6mm}
	\label{transition}
\end{figure}



\vspace{-2mm}
\subsection{Bypass Leader Replication Reply}\label{design_bypass leader}

As in leaderless replication methods~\cite{tapir, janus, mdcc, carousel, ov, starry}, they make the replication bypass leader to reduce commit latency. We also propose the bypass leader replication reply method by utilizing \coss.
In this approach, replication replies are directed to bypass the leader, while the transaction order is still maintained by the leader. By doing so, the commit latency can be reduced without compromising the conflict resolution efficiency. 



Firstly, we clarify that the \co that co-locates with the client is the \emph{\primary}, which is the final commit decision maker of the transaction initiated by this client.
As depicted in Fig.~\ref{overview}(b), after the leader replicates to followers, each follower notifies the co-located \co about the vote and includes the replication reply in the message (message \ding{174}). Subsequently, the co-located \co transfers the replication reply directly to the \primary (message \ding{175}). 
In this way, the replication reply can bypass the leader, which reduces one \cross network delay for the \primary to be aware of the replication result. Therefore, after 1.5 \cross RTTs, the \primary has received the replies from each participant's majority of followers, which indicates that replication on each participant shard has been completed. 



\subsection{Transition of Transaction State} \label{state transition}
Fig.~\ref{transition} illustrates the state transitions of transactions. The main difference from 2PC is the introduction of the \precommit state. After the \co collects votes from the co-located replicas, the transaction is set to \precommit state (arrow \ding{174}). When the \primary receives replication replies from a majority of replicas in each shard, signifying successful replication, it transitions the transaction to the \commit state (arrow \ding{176}). One exception is that if a majority of replicas in a participant shard fail, the replication cannot be finished, in such a case, even if a transaction is \precommit, it has to be aborted (arrow \ding{178}).

\subsection{Architecture}\label{design_architecture}

\begin{figure}[t]
	\centerline{\includegraphics[width=6cm]{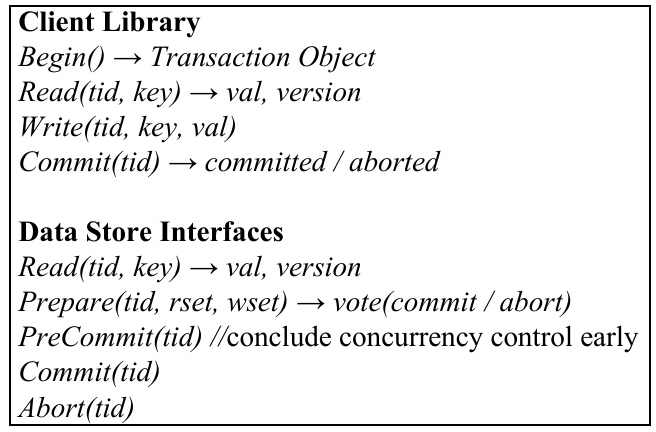}}
	\vspace{-2mm}
	\caption{Interfaces in \ourwork.}
	\label{interface}
 \vspace{-6mm}
\end{figure}

\noindent \textbf{Server Setup.}\label{servers} As the backend for applications with global users, the \gddb database is deployed in multiple datacenters across regions. Specifically, the database is partitioned into \emph{data shards}, with 2\majority + 1 replicas per shard, each located in a distinct datacenter. 

\emph{\Cos} are distributed servers located in all datacenters that make \precommit decisions in a decentralized manner. For each individual transaction, there is a \textit{\primary}, which is the \co located in the same datacenter as the client. The \primary is responsible for making the final commit decision, since it can ensure the replication is completed. Other \coss act as votes collectors and \precommit decision notifiers, facilitating the early termination of \cc for participant leaders.



\noindent \textbf{Client Library.}\label{client} \zzh{Clients in \ourwork are application servers that run within the same datacenter as database replicas. They interact with the database through the interfaces provided by the client library, illustrated in Fig.~\ref{interface}. When starting a transaction, the client invokes the \textit{Begin()} function, generating a transaction object with a unique transaction ID (\textit{tid}). The transaction ID is designed as a tuple comprising the client ID and the transaction counter of the client. After creating the transaction, the client calls the \textit{Read(tid, key)} function to read data, either from the shard leader or the local replica (details in \S ~\ref{optimization}), and buffers the results locally. For write operations, the client uses \textit{Write(key, val)} function to perform writes. The write values are also buffered locally until the client invokes the \textit{Commit(tid)} function to send the commit request to the database.}

\noindent \textbf{Data Store.}\label{store}
The implementation of \ourwork is agnostic to the underlying data store. For the sake of simplicity, we use the key-value data store. 
Each shard replica contains a key-value data store responsible for data storage and concurrency control. The data store provides several interfaces for transactions as shown in Fig.~\ref{interface}. On receiving the commit request from the client, the data store invokes the \textit{Prepare(tid, wset, rset)} function to validate whether the transaction can be committed. Then it starts replication to send the transaction log and its vote to all datacenters.

Once a shard leader receives the \precommit decision from its co-located \co, it invokes the \textit{PreCommit(tid)} function to conclude the \cc. Upon receiving the final commit result from the \primary, the shard leader will invoke either the \textit{Commit(tid)} or \textit{Abort(tid)} function to end the transaction accordingly.

\section{\ourwork Protocol}\label{protocol}
\subsection{Commit Processes in \ourwork}\label{fast_commit}


After a new transaction is initiated, the client executes the transaction logic and generates the read and write sets as inputs for \ourwork. During execution, read operations are performed by invoking the \emph{Read(tid, key)} function. Usually, the read request is handled by the participant leader to obtain the most up-to-date value, and subsequently a new entry $\langle$ \emph{key, version} $\rangle$ is added to the read set. For write operations, the new values are also stored in the write set as $\langle$ \emph{key, new\_value} $\rangle$. Once the transaction is \executed, the \emph{Commit(tid)} function is invoked and the transaction enters the PreCommit phase of the \ourwork protocol.



\begin{figure}[t]
	\centerline{\includegraphics[width=8cm]{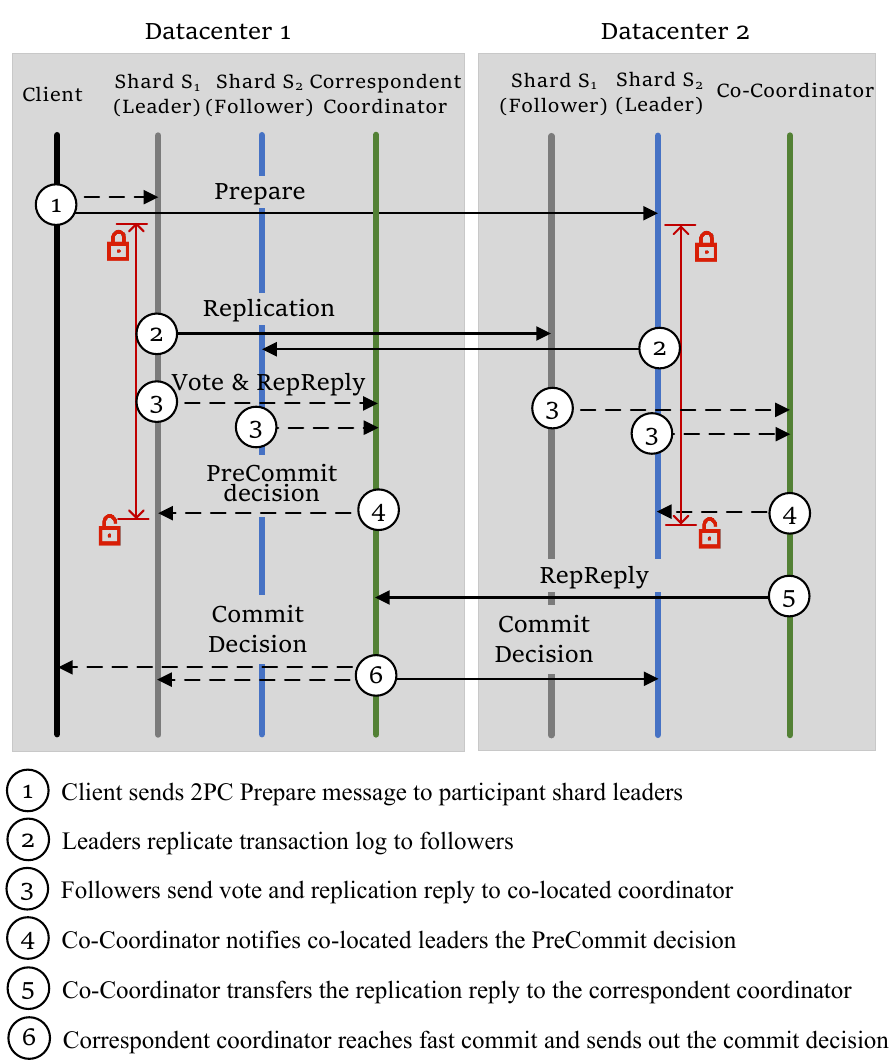}}
	\vspace{-3mm}
	\caption{Example of committing transaction with the \ourwork protocol. Solid and dashed arrows stand for inter-datacenter and intra-datacenter messages, respectively.}
	\vspace{-5mm}
	\label{d2pc}
\end{figure}

\vspace{1mm}
\noindent \textbf{PreCommit Phase.} When the client invokes the \emph{Commit} function, a \preparereq request is sent to all participant leaders (Process \ding{172} in Fig.~\ref{d2pc}). Each participant leader receiving the \preparereq request retrieves the transaction ID (tid) along with the transaction's read and write set specific to that shard. Subsequently, the participant leader invokes the \emph{Prepare} function of the data store to validate whether the transaction can be successfully committed. \zzh{If the participant leader agrees to commit the transaction, it is marked as \preparestate.} The leader then generates a log entry for the transaction, which includes tid, read and write sets. \zzh{The log entry, along with the vote and involved shard list, is subsequently replicated to all replicas, including the leader itself (Process \ding{173}). In cases where a datacenter lacks a replica, the leader directly sends its vote to the \co of that datacenter.}


\zzh{Upon receiving the replication message, each replica notifies the co-located \co of the vote, its replication reply, and the involved shard list (Process \ding{174}).} At this point, each \co can collect votes of each shard and make the \precommit decision (Process \ding{175}). After that, each \co forwards the replication reply to the \primary (Process \ding{176}). Once the \primary receives replication replies from the majority of replicas for each participant, it confirms the fault-tolerant state of all participant shards. At this point, the \primary can safely commit the transaction and the transaction enters the \emph{Commit} phase.


The process described above, where the \primary directly obtains the replication reply bypassing the leader, can be considered the fast path. \zzh{Additionally, there is a slow path that acts the same as commit processes in \layered as described in Fig.~\ref{2pc}. Since both fast and slow paths make commit decision according to the vote of each shard, they always produce the same output.} This slow path is retained to ensure that the \primary can still learn the replication results in case half of the \coss are failed and it cannot obtain the replication replies in a bypass leader way. In the slow path, the \primary will directly transform the state from \preparestate to \commitstate (as shown by arrow \ding{197} in Fig.~\ref{transition}).

\vspace{1mm}

\noindent \textbf{Commit Phase.} 
Once the \primary has made the decision to either commit or abort, it promptly sends a response to the client (Process \ding{177}). Subsequently, it asynchronously notifies all participant leaders of the decision. This notification step is not part of the commit path and does not impact the commit latency. 

\vspace{1mm}

\noindent \textbf{Commit Latency Analysis. } 
The transaction commit process begins when the client sends a \preparereq message and ends upon receiving a notification from the \primary. As the notification from the \primary is an intra-datacenter communication, which is trivial compared to the delay required for \cross communication, it can be assumed that the transaction commit ends when the \primary makes the final commit decision. 

The \primary must meet two conditions to make the commit decision. ($i$) It must receive votes from all participant shards. The message path is client $\stackrel{inter}{\rightarrow}$ shard leader $\stackrel{inter}{\rightarrow}$ co-located follower of the \primary $\stackrel{intra}{\dashrightarrow}$ \primary. As the message from co-located follower to the \primary is intra-datacenter, the total message delay is one round of \cross communication. ($ii$) It must receive the replication replies from the majority of followers of each shard. The message path is client $\stackrel{inter}{\rightarrow}$ shard leader $\stackrel{inter}{\rightarrow}$ shard follower $\stackrel{intra}{\dashrightarrow}$ co-located \co $\stackrel{inter}{\rightarrow}$ \primary. As the message from the shard follower to its co-located \co in intra-datecenter, it requires a total of 1.5 rounds of \cross communication. Therefore, the overall commit latency is 1.5 \cross RTTs.

\zzh{An important observation to highlight is that in the commonly adopted three-replica deployment, if a replica co-locates with the correspondent coordinator, the commit latency is only 1 \cross RTT.} In this deployment, two replicas already form a majority. Consequently, besides the leader, the \primary only needs to wait for replication reply from the follower co-located with it. The message path is client $\stackrel{inter}{\rightarrow}$ shard leader $\stackrel{inter}{\rightarrow}$ co-located follower $\stackrel{intra}{\dashrightarrow}$ \primary. Since the communication from the co-located follower to the \primary is intra-datacenter, it only takes 1 \cross RTT to complete the commit process.

\vspace{1mm}

\vspace{-2mm}
\subsection{Decentralized Commit via \Cos}\label{shorten_path}



As shown in Fig.~\ref{d2pc}, upon collecting votes of all participant shards from local followers, the \co independently makes the \precommit decision. Precisely, if \abort exists in the votes, the transaction will be aborted. Otherwise, the transaction will be precommitted. Since each \co can only make the decision according to the votes of each participant, consistent \precommit decisions are always reached. After they make the decision, they will notify the co-located participant leaders immediately.


When a participant leader receives the \precommit decision, it promptly concludes the \cc, e.g., releases locks in case of 2PL and removes the transaction from the validation list in OCC.
Then transaction updates are visible. However, to maintain database recoverability, it is essential to track read-after-write dependencies, as detailed in \S~\ref{recoverability}.


\zzh{By leveraging decentralized commit, the \cc starts when the participant leader receives the \prepare message, and concludes at the time of receiving the \precommit decision. The message path is shard leader $\stackrel{inter}{\rightarrow}$ shard followers $\stackrel{intra}{\dashrightarrow}$ co-located \co $\stackrel{intra}{\dashrightarrow}$ co-located shard leader. Since the path contains only one \cross message, the \cc length is 0.5 \cross RTT. 

Note that when 2PL is adopted, the \cc starts when read locks are acquired during execution, which will extend the \cc length. But with the read optimization that will be introduced in \S~\ref{optimization}, the \cc length of \dlock can also be reduced to only 0.5 \cross RTT. }

After a transaction is committed, \coss establish consensus to ensure the fault tolerance of the commit decision. For each individual transaction, the \primary is responsible for replicating the commit decision to a majority of \coss. Ensuring the fault tolerance of the commit decision simplifies the resolution of 2PC blocking. As outlined in \S~\ref{fault-tolerance}, in the event of a \primary failure, the commit decision can be recovered from other \coss.

To achieve this, once the \primary makes the final commit decision, it notifies all \coss. Each \co will end the processing of this transaction and reply to the \primary. When the \primary receives replies from a majority of \coss, the commit decision reaches fault-tolerant. It is important to note that the replication of commit decisions is independent of the commit of transactions, which means it is not on the transaction commit path and has no impact on the commit latency.

\zzh{
Since each \co handles all transactions, to prevent \coss from becoming bottlenecks under increasing loads, we design a \co sharding strategy that shards \coss into multiple \co groups. For example, when \coss become the bottleneck, they can be partitioned into N groups, with N co-coordinators in each datacenter. In this way, the load on one co-coordinator can be balanced on N co-coordinators by taking the transaction ID modulo N. }

\vspace{-2mm}
\subsection{Dependency Tracking}\label{recoverability}
If the \cc is concluded before the transaction is committed, it is crucial to have strategies to ensure both \emph{serializability} and \emph{recoverability}. To ensure serializability, all dependencies, including \war, \raw, and \waw dependencies, should be tracked to prevent them from forming cycles. To ensure recoverability, we should prevent a transaction that reads uncommitted data from being committed. For example, $T_{2}$ reads the uncommitted updates of $T_{1}$ and commits before $T_{1}$. If $T_{1}$ is later aborted and the server failure occurs, $T_{2}$ cannot be recovered as it has read non-existent data. Hence, to ensure recoverability, \raw dependencies should be tracked. In \ourwork, we choose to conclude the \cc when the \precommit decision is made. At this stage, the global serial order has been established. Therefore, concluding the \cc will not violate serializability. Thus, \ourwork only needs to track \raw dependencies.

\ourwork introduces a \precommit list for each tuple, which can efficiently identify and manage \raw dependencies. When a participant leader receives the \precommit decision of a transaction, it concludes the \cc and adds the transaction id to the \precommit list of each write key. For a subsequent transaction that accesses a specific key, it checks the key's \precommit list to determine if any transaction has completed the \cc but has not committed yet. If so, a \raw dependency is identified. Note that the updates made by the \precommit transaction do not take effect in place. Instead, the new values of the updated keys are maintained in the write set in memory. Therefore, when a new transaction reads the key, it locates the last \precommit transaction and uses its transaction ID to retrieve the write set to obtain the most up-to-date value.
This approach simplifies the process of aborting since there is no need to maintain the undo log. When a transaction is aborted, \ourwork can directly remove it, along with all transactions dependent on it, from the \precommit list.

\LinesNumbered
\begin{algorithm}[t]
	\caption{Function calls in \ourwork}\label{functions}
	\tcc{$tuple.precommit$ \emph{\# List\ of\ \precommit\ transactions\ on\ the\ tuple}}
    \SetKwBlock{Function}{\underline{Function} Read($txn, tuple$)}{end}
	\Function(){
        $t \gets the\ last\ entry\ in\ tuple.precommit$

        $t.out.add(txn)$

        $txn.in++$
	}
 \tcp{\emph{calls when receiving the \precommit decision from the co-located \co.}}
	\SetKwBlock{Function}{\underline{Function}
	PreCommit($txn$)}{end}
	\Function(){
	\For{$\forall \ tuple \in txn.wset$}{

            
            $tuple.precommit.add(txn)$\
		}
	}
 \tcp{\emph{calls when receiving the commit decision from the \primary.}}
 \SetKwBlock{Function}{\underline{Function}
	Commit($txn, decision$)}{end}
	\Function(){
    \For{$\forall \ t \in txn.out$}{
        \If{$decision == \commitstate$}{
            $t.in--$
        }
        \ElseIf{$decision == \abortstate$}{ 
            $t.in \gets -1 $
        }
    }
	\For{$\forall \ tuple \in txn.wset$}{
            $tuple.precommit.remove(txn)$\
		}
	}
\end{algorithm}

After identifying \raw dependencies, \ourwork effectively manages them using the register-and-report method~\cite{hakaton2}. On each participant shard, each transaction maintains a counter called \emph{in}, which records the number of transactions it depends on (i.e., the uncommitted transactions whose updates are read by it). Additionally, each transaction maintains a list named \emph{out}, which captures all the transactions that depend on it (i.e., the transactions that have read its updates).

The maintenance of \raw dependencies is illustrated in Algorithm~\ref{functions}. For each read operation, the system checks if the read tuple has any \precommit transactions. If so, a \raw dependency is detected, and the corresponding \emph{in} and \emph{out} counters are updated accordingly (lines 2-4). Upon receiving the \precommit decision, the participant leader invokes the \textit{PreCommit} function to add the transaction to the \precommit list of all write keys (lines 6-7). When the final commit message is received, the \textit{Commit} function is called, which removes the transaction from \precommit lists (line 15). \zzh{If the transaction is committed, all transactions that depend on it decrement their \emph{in} counters by one. Conversely, if the transaction is aborted due to replication failure (arrow \ding{198} in Fig.~\ref{transition}), the \emph{in} counter of the dependent transaction is set to -1 (lines 9-13).}

To ensure that transactions are committed in accordance with the \raw dependency order, we enforce a rule that a participant leader cannot vote \commitstate for a transaction if its \emph{in} counter is greater than 0. Only when all its dependent transactions are committed, can the transaction be committed. In the case where a transaction's \emph{in} counter is set to -1, indicating that it will be cascade aborted, the participant leader directly invokes the \textit{Commit} function with the decision of \abortstate. This ensures the proper commit order of transactions based on their dependency relationships.


\subsection{Failure and Recovery}\label{fault-tolerance}


\noindent \textbf{\PRIMARY Failure.} The \primary is the only node that can make the final commit decision, and as such, its failure may cause participant leaders and other \coss to time out waiting for the decision. We will discuss how to handle \primary failure in different cases.

\textbf{Case 1.} If the \primary fails after the final decision has been sent to some participants and \coss but not all of them, some \coss will timeout while awaiting the decision. In such a scenario, a \co is elected as the \primary and initiates the recovery phase by asking other \coss to determine if anyone has received the decision. Upon receiving a reply containing the decision, the decision will be accepted and notified to participant leaders and other \coss.

\textbf{Case 2.} 
In the event that the \primary fails before the final decision is sent out, a \co is elected as the \primary and initiates the recovery phase as distributed in \textbf{Case 1}. Since the decision was not made or sent out before the previous \primary failed, no \co has received the decision. Consequently, this \co proceeds with the termination protocol, which involves communicating with all participant shard leaders to obtain their votes and replication results. Once the replies are received from all participant shards, this \co makes the commit decision. It then notifies all participant leaders and \coss, and concludes the process of the undetermined transaction.



\vspace{1mm}

\noindent \textbf{\Co Failure.} If a \co fails, the participant leader co-located with it may not receive the \precommit decision. When the leader eventually receives the decision from the \primary, it can safely end the transaction. If more than \majority \coss fail, there is a possibility that the \primary may not receive the replication replies through the fast path. Nevertheless, the \primary can still obtain the replication results from participant leaders via the slow path, as described in \S~\ref{fast_commit}. Thus, the failure of a \co does not bother the successful commit of transactions.

\vspace{1mm}
\noindent \textbf{Participant Shard Replica Failure.}
When a participant shard leader fails, the \primary can still receive the vote and replication replies from shard followers. Once the \primary has collected votes and replication replies from \majority+ 1 followers, it can ensure that the replication has completed and makes the final commit decision. If less than \majority+ 1 replies are received, the \primary cannot be certain whether the replication has been successful. In this scenario, the \primary will wait for a new leader to be elected and process this transaction. Once the new leader notifies the replication result, the \primary can make the final decision and conclude the recovery of this transaction. 

If a shard follower fails, the co-located \co may not receive the vote of this shard. This prevents participant leaders from concluding \cc early because the co-located \co cannot make the \precommit decision. Therefore, only after receiving the final commit decision from the \primary, the participant leaders in this datacenter can safely end the transaction.

\subsection{Read Optimization}\label{optimization}

\begin{table}[t]
	\centering
	\caption{\bf Network latency between datacenters (ms). Datacenters are located in Hangzhou (South China), Beijing (North China), San Francisco (US West), Virginia (US East), and Frankfurt (Europe).}
	\label{network latency}
 \resizebox{1\columnwidth}{!}{
	\begin{tabular}{c|ccccc}
		& \textbf{\makecell[c]{Hangzhou}} & \textbf{\makecell[c]{Beijing}} & \textbf{\makecell[c]{San Francisco}} & \textbf{\makecell[c]{Virginia}} & \textbf{\makecell[c]{Frankfurt}} \\
		\hline
		\textbf{\makecell[c]{Hangzhou }} & 0.2 & 30 &  140 & 203 & 231\\
		\textbf{\makecell[c]{Beijing}}  &  & 0.2 & 150 & 215 & 240\\
        \textbf{\makecell[c]{San Francisco}} & & & 0.2 & 67 & 151\\
        \textbf{\makecell[c]{Virginia}} & & & & 0.3 & 98\\
		\textbf{\makecell[c]{Frankfurt}} & & &  &  & 0.25\\
		\vspace{-4mm}
	\end{tabular}
 }
\end{table}

During transaction execution, read operations are typically served by each shard's leader to ensure data recency. In practice, read operations constitute a large portion of the workload, and accessing the remote leader in a different datacenter incurs substantial costs. Considering that each participant shard may have a replica that co-locates with the client, we optimize read operations by enabling the client to directly read from the local replicas, thus reducing the need for \cross communication.

However, reading from follower replicas may read stale data. To ensure serializability, we will perform a read version verification during committing. The \prepare message will be accompanied by a read set containing the read data and the read versions of the transaction. When a leader receives the \prepare message, it first verifies if read versions are up-to-date, and then aborts those who have read stale data. Only when all read operations pass the recency validation, the leader continues to process the transaction. 


The read optimization strategy is compatible with both 2PL and OCC. Since OCC is a verification-based approach, integrating the read optimization aligns naturally with OCC. However, in 2PL, acquiring read locks during execution is crucial for fairness and preventing starvation. Introducing read optimization in 2PL may cause starvation, as large read transactions could frequently abort due to stale reads without locks during execution, which can be avoided in standard 2PL protocol. To address this, when transactions are aborted due to stale read, a second execution is enforced to acquire read locks from the shard leader. Concurrently, pending transactions will receive locks based on their timestamp order. This allows \ourwork to mitigate performance penalties associated with remote reads during execution while preserving fairness in 2PL. In our evaluation, detailed in \S~\ref{evaluation}, we separately evaluate the performance of 2PL with and without read optimization, whereas OCC consistently incorporates read optimization.


\subsection{Integrating into the \LAYERED Databases}
\ourwork introduces minor modifications to the commit process of the \layered \gddb database. As mentioned in \S~\ref{fast_commit}, \ourwork retains a slow path identical to the traditional commit process of the \layered database. While in the fast commit path, the only modification to database servers (i.e., data replicas) is that each server should send shard's vote and replication reply to the co-located \co, introducing only an \intra message. Therefore, integrating \ourwork into the \layered \gddb databases necessitates minimal changes to the original commit protocol, incurring only a marginal cost on each database server. Actually, most of the work in \ourwork is carried out by the independent \co group.
To facilitate \cc shortening, the data store should be designed for tracing \raw dependencies. This design allows the shard leader to conclude \cc early for higher concurrency.
\section{Evaluation}\label{evaluation}
\subsection{Experimental Setup}\label{setup}

\noindent \textbf{The Testbeds.} 
All experiments were conducted on Alibaba Cloud ECS instances distributed across five datacenters. Table \ref{network latency} outlines the network latencies between datacenters. By default, the experiments were conducted with a 3-replica deployment, where the replicas were located in Hangzhou, San Francisco, and Frankfurt. For the 5-replica deployment, the replicas in Beijing and Virginia were also involved.

\begin{table}[t]
	\centering
	\caption{\bf Transaction profile for Retwis workload.}
	\label{retwisworkload}
 \resizebox{0.8\columnwidth}{!}{
	\begin{tabular}{cccccccc}
		\toprule[1pt]
		\textbf{Transaction Type} & \# \textbf{gets} & \# \textbf{puts} & \textbf{workload\%} &\\
		\midrule
		Add User & 1 & 3 & 5\% &\\
		Follow/Unfollow & 2 & 2 & 15\%\\
		Post Tweet & 3 & 5 & 30\%\\
		Load Timeline & rand(1,10) & 0 & 50\%\\
		\bottomrule[1pt]
   \vspace{-6mm}
	\end{tabular}
 }
\end{table}

Each cloud server was equipped with 4 virtual CPU cores and 8GB of memory. 
The system configuration involved 3 data shards, with each shard consisting of 3 - 5 replicas. Each cloud server hosted a replica for each shard. The leaders of shards were distributed among different datacenters. Furthermore, within each datacenter, a server acted as the \co to serve co-located servers.

\vspace{1mm}
\noindent \textbf{Candidates for Comparative Study.} \zzh{First, we compared \ourwork with 2PC-based methods. Regarding concurrency control methods, \ourwork can be combined with both 2PL and OCC. Thus we conducted a comparison between \ourwork and the standard transaction protocol like Spanner~\cite{spanner1, spanner2} named \lock, which is the combination of 2PC, 2PL, and Multi-paxos. There is also an approach named \occ, which can be seen as Spanner's implementation with OCC as the concurrency control mechanism. Both \lock and \occ represent \layered architecture, as illustrated in Fig.~\ref{2pc}.}  Notably, implementations were optimized to eliminate the replication of commit decision (Process \ding{175} in Fig.~\ref{2pc}). Therefore, both \lock and \occ exhibit a commit latency and \cc length of 2 \cross RTTs.




\begin{figure*}[t]
	\vspace{-6mm}
	\centerline{\includegraphics[width=12cm]{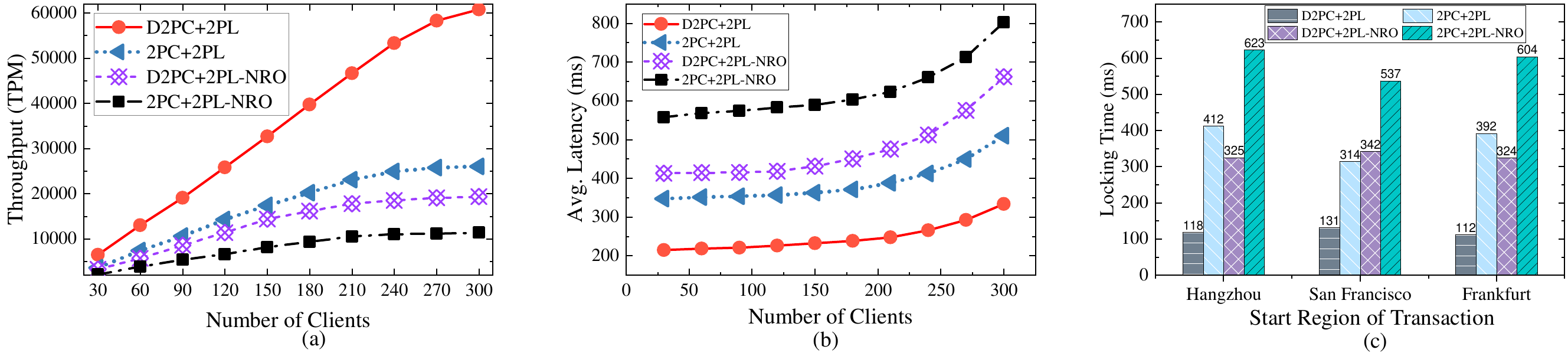}}
	\vspace{-4mm}
	\caption{Performance under different clients number (2PL).}
	\vspace{-2mm}
	\label{2pl clients}
\end{figure*}

\begin{figure*}[t]
	\centerline{\includegraphics[width=16cm]{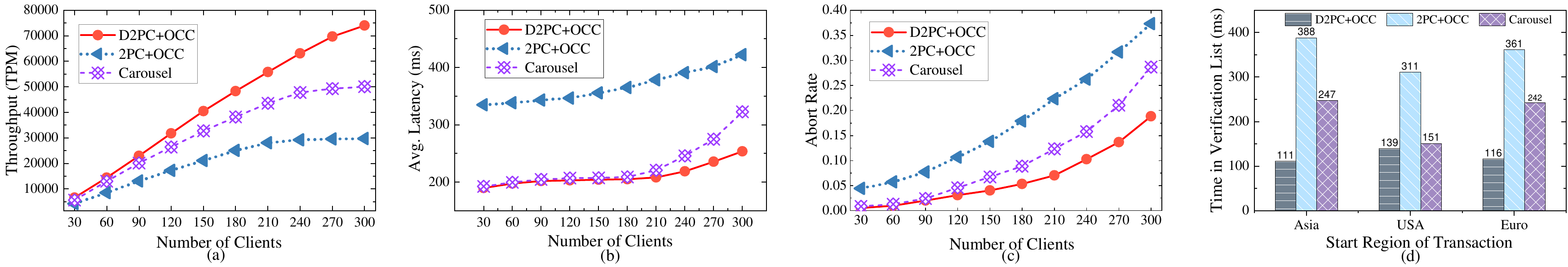}}
	\vspace{-4mm}
	\caption{Performance under different clients number (OCC).}
	\vspace{-2mm}
	\label{occ clients}
\end{figure*}

\begin{figure*}[t]
	\centerline{\includegraphics[width=16cm]{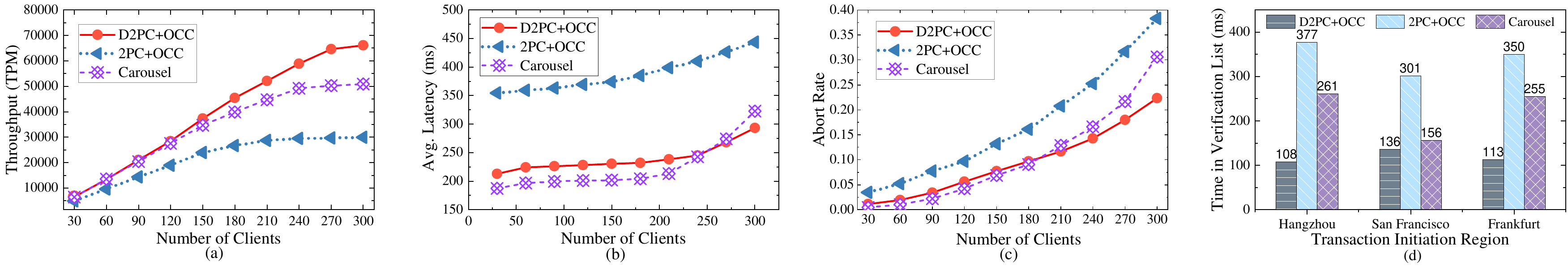}}
	\vspace{-4mm}
	\caption{Performance under 5-replica deployment (OCC).}
	\vspace{-4mm}
	\label{replicas}
\end{figure*}

Additionally, we conducted a comparison between \ourwork and Carousel~\cite{carousel}, which represents the transaction commit approach of the \integrate. In conflict-free scenarios, Carousel allows each replica to make the commit decision and reach a consensus in one round of \cross communication. However, in conflict scenarios, it falls back to a slower path where the leader of each shard makes the commit decision, requiring at least two rounds of \cross communication. \zzh{Regarding the concurrency control protocol, Carousel utilizes the OCC. Carousel also utilizes a read optimization technique in its paper, similar to the strategy outlined in \S~\ref{optimization}.}


To ensure a fair comparison, we implemented all approaches using the same code prototype. Additionally, all methodologies defaultly integrated the read optimization technique detailed in \S~\ref{optimization}. For 2PL, we also conducted evaluations without read optimization, denoted as \locknro and \ourworknro.


\vspace{1mm}

\noindent \textbf{Workload.} We used two workloads for evaluation. The first was a synthetic workload for the Retwis application, which simulates Twitter's functionality. The Retwis workload contains four types of transactions, as outlined in Table \ref{retwisworkload}. On average, each transaction accesses 4-10 data items across 2-3 shards. The second workload was YCSB+T~\cite{ycsbt}, an extension of YCSB~\cite{ycsb} that supports transactions. We selected these workloads for consistency and comparability, as they were also employed in the evaluation of Carousel\cite{carousel}.

\vspace{-2mm}
\subsection{Performance with Varied Loads}\label{scalability}

\begin{figure*}[t]
	\centerline{\includegraphics[width=12cm]{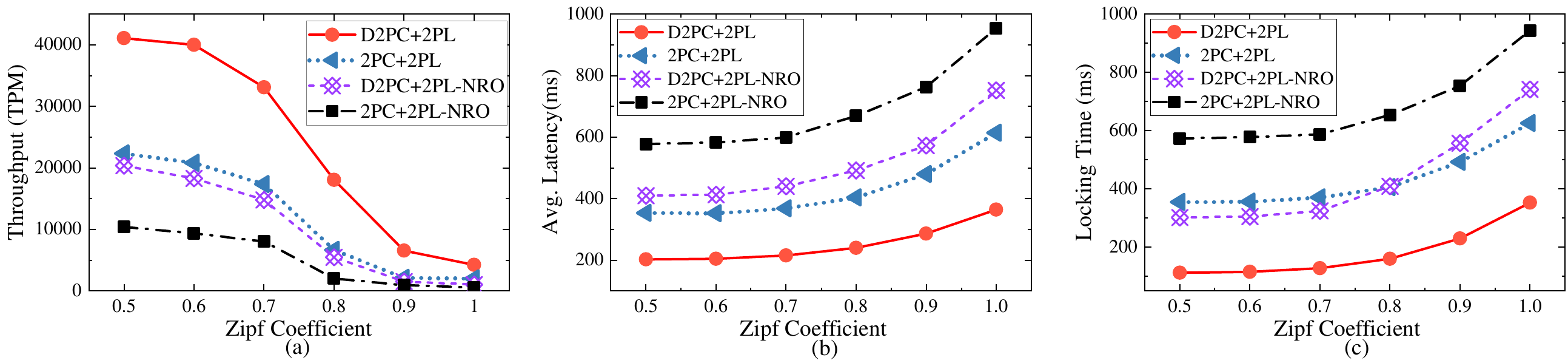}}
	\vspace{-5mm}
	\caption{Performance under contention (2PL).}
	\vspace{-3mm}
	\label{2pl skew}
\end{figure*}

\begin{figure*}[t]
	\vspace{-2mm}
	\centerline{\includegraphics[width=16cm]{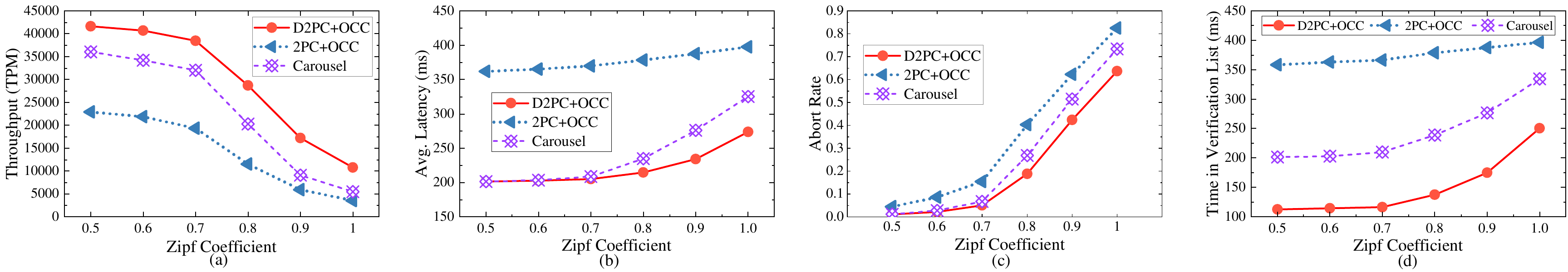}}
	\vspace{-5mm}
	\caption{Performance under contention (OCC).}
	\vspace{-4mm}
	\label{occ skew}
\end{figure*}

\vspace{1mm}

Firstly, we evaluated the performance of different approaches under the regular Retwis workload with medium contention (Zipf coefficient = 0.7).


\zzh{Fig.~\ref{2pl clients} illustrates the performance with 2PL. We first set both \dlock and \lock disable the read optimization to clearly show the improvement brought by \ourwork. We can see that \ourworknro achieves 1.73 $\times$ throughput than \locknro. This is because \ourworknro can reduce the latency by 25\% and shorten the \cc length by 44\%. After enabling read optimization, the remote reading during execution is eliminated, therefore both \dlock and \lock achieve significant improvement in throughput and latency. Enabling read optimization further expands the performance advantages of \dlock. As shown in Fig.~\ref{2pl clients}(a), (b), and (c), compared to \lock, \dlock can realize a 2.33 $\times$ improvement in throughput, 42\% reduction in latency, and 66\% reduction in locking time. }


\zzh{
Fig.~\ref{occ clients} illustrates the performance under OCC. \docc demonstrates a substantial performance improvement compared to the other approaches. As the number of clients increases, the throughput of \docc exhibits a more rapid growth rate compared \occ. 
The enhanced performance of \docc can be attributed to low latency and short \cc length. 
For example, with 300 clients, the commit latency of \docc is approximately 250 milliseconds, which is reduced by about 43\% compared to \occ, and \docc also achieves a reduction in the \cc length of 64\%, enabling higher concurrency and resulting in the throughput increases to 2.43 $\times$ that of \occ.}

We also compared \ourwork with Carousel, an \integrate approach that can achieve fast committing. In comparison, \docc achieves similar or lower commit latency. As depicted in Fig.~\ref{occ clients} (b), when the load is low, the latency of \docc and Carousel is comparable, because both of them can commit a transaction in one \cross RTT. As the client number increases, the likelihood of conflicts rises. In such a scenario, Carousel's fast commit path fails and switches to the slow path, where transactions go through two \cross RTTs to be committed. Therefore, \docc outperforms Carousel in transaction latency as the concurrency increases. In terms of \cc length, \docc demonstrates a significant advantage. While the \cc length of Carousel is one \cross RTT, \docc is only 0.5. This reduced \cc length contributes to \docc's higher throughput. As depicted in Fig.~\ref{occ clients}, under 300 clients,
\docc achieves a 47\% higher throughput than Carousel, with a 23\% reduction in latency and a 46\% reduction in \cc length.
Meanwhile, the short \cc leads to fewer conflicts, resulting in a 34\% reduction in abort rates.

\vspace{-2mm}
\subsection{Peformance under 5-replica Deployment}\label{replica}
We further measured the performance using a 5-replica deployment, with replicas located across all five datacenters. The concurrency control protocol was OCC, and the workload was Retwis with Zipf coefficient set to 0.7.

Firstly, we observe that \docc still shows a significant reduction in \cc length (Fig.~\ref{replicas}(d)), and achieves the highest throughput (Fig.~\ref{replicas}(a)), which is 1.29 $\times$ that of Carousel and 2.2 $\times$ that of \occ.

As for latency, given that the initial 3-replica deployment was already widely geographically distributed across South China, US West, and Europe, the inclusion of two additional replicas in the 5-replica deployment does not lead to a significant increase in latency for \occ and Carousel. 
For \docc, the commit latency slightly increases in the 5-replica deployment due to the requirement of 1.5 \cross RTTs to complete the commit. Nevertheless, \docc still achieves a similar commit latency to Carousel and significantly lower latency than \occ. When the number of clients exceeds 240, \docc exhibits a lower abort rate and average latency compared to Carousel. 
This distinction becomes more pronounced under high concurrency scenarios, such as under 300 clients, the latency of \docc is lower than that of Carousel due to Carousel's need for additional communication to resolve conflicts.

\vspace{-2mm}
\subsection{Performance under Contention}\label{contention}
By varying the Zipf coefficient, we simulated different contention levels and evaluated the performance of each approach. The workload adopted was Retwis, and the client number was fixed at 150.

Fig.~\ref{2pl skew} and \ref{occ skew} depict the performance under contention with 2PL and OCC respectively. As the contention level increases, the throughput of all approaches gradually decreases and shows a sharp decline trend when the Zipf coefficient exceeds 0.7. 
Comparatively, the throughput of \ourwork remains significantly higher than other approaches in both two concurrency control protocols under high contention loads. For example, as shown in Fig.~\ref{occ skew}, when using OCC and setting Zipf coefficient to 0.9, the throughput of \ourwork surpasses \occ by 2.35 $\times$. Compared with Carousel, \ourwork consistently achieves higher throughput, surpassing it by 1.41 $\times$. Similarly, in the case of 2PL, Fig.~\ref{2pl skew} shows that \ourwork outperforms \lock with a throughput that is at most 3.06 $\times$ higher. When read optimization is disabled, \ourworknro also achieves 1.8 $\times$ throughput than \locknro. The performance advantages of \ourwork can be attributed to its shorter \cc, as evident from Fig.~\ref{2pl skew} (c) and \ref{occ skew} (d). Regardless of the contention load, \ourwork consistently exhibits the shortest \cc.
This reduction in \cc mitigates contention, resulting in the lowest transaction abort rate, as demonstrated in Fig.~\ref{occ skew} (c).




\begin{figure}[t]
 \hspace{-8mm}
	\centerline{\includegraphics[width=8cm]{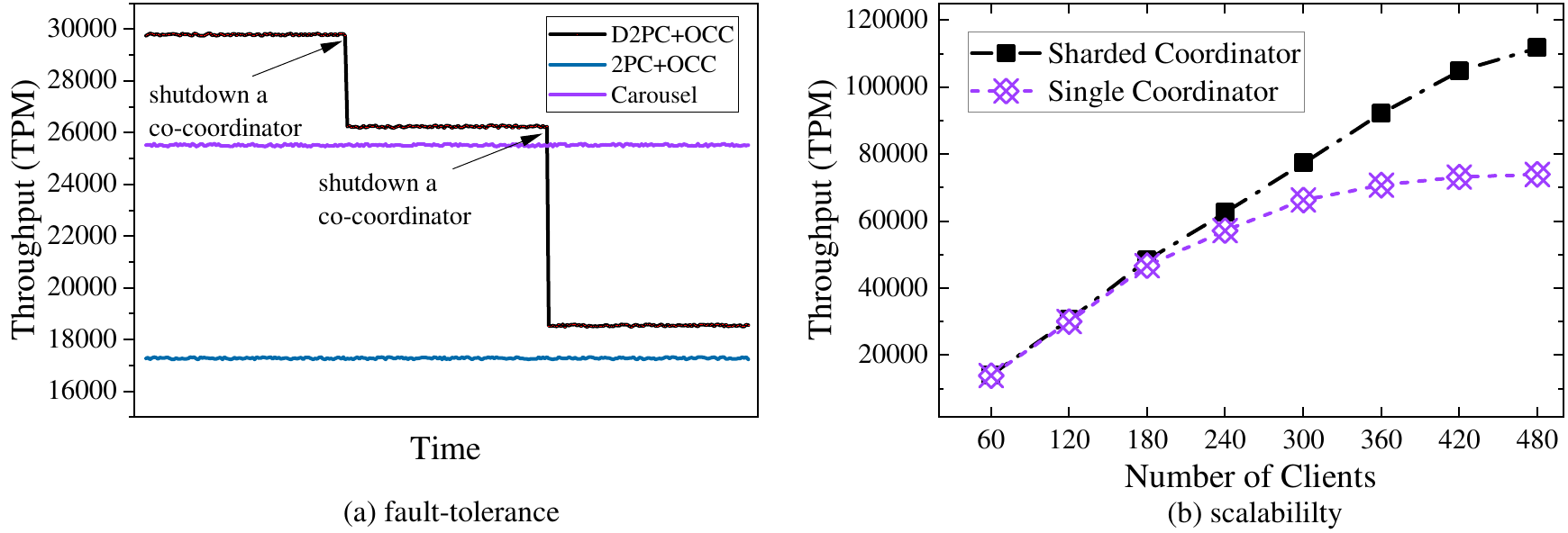}}
	\vspace{-4mm}
	\caption{Fault-tolerance and scalability of \coss.}
	\vspace{-2mm}
	\label{coors}
\end{figure}

\begin{figure*}[t]
\vspace{-4mm}
\centering
\hspace{10mm}
	\begin{minipage}[t]{0.24\textwidth}
		\centering
		\includegraphics[width=3.7cm]{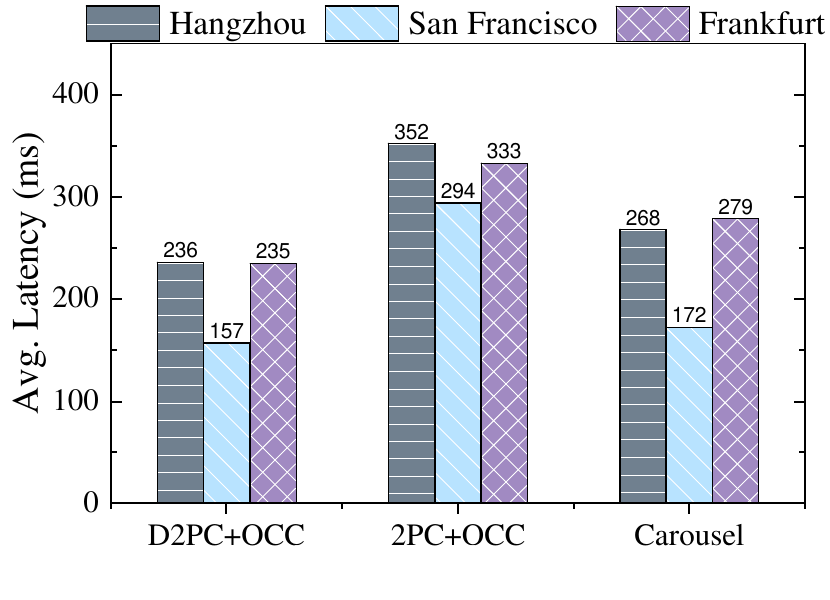}
		\vspace{-6mm}
		\caption{Comparsion of commit latency}\label{latency}
	\end{minipage}
	\hspace{-9mm}
	\begin{minipage}[t]{0.74\textwidth}
		\centering
		\includegraphics[width=12cm]{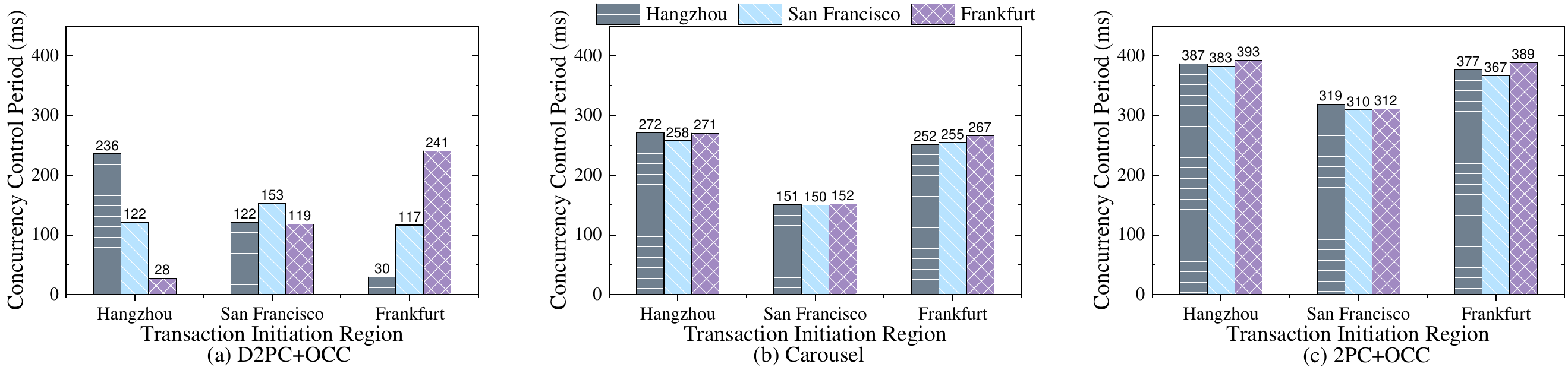}
		\vspace{-4mm}
		\caption{Comparsion of the length of \cc}\label{cc}
	\end{minipage}
      
	\vspace{-4mm}
\end{figure*}
\vspace{-2mm}
\zzh{\subsection{Fault-tolerance and Scalability of \Cos}\label{scalability of co}
In this experiment, we first evaluated the throughput of \ourwork under \co failures. In Fig.~\ref{coors}(a), we artificially shut down two \co servers. Since \co failure results in some transactions failing to be PreCommitted, the throughput of \ourwork will decrease. When only one \co is alive, \ourwork essentially behaves like \occ, requiring two rounds of communication to commit a transaction, leading to lower throughput than Carousel. However, since participant leaders co-located with the alive \co can still receive \precommit decision and conclude the \cc, \ourwork maintains an advantage over \occ when there is only one \co.

We further evaluated the efficacy of the \co sharding strategy in improving the scalability of \coss. The workload adopted was Retwis, and the Zipf coefficient was 0.7.
To simulate the scenario where \coss could bottleneck performance, we allocated only 1 CPU and 512 MB memory to \co servers. As shown in Fig.~\ref{coors}(b), when a single \co group is deployed, throughput stabilizes as the client number increases. After sharding \coss into two groups (with two \coss in each datacenter), the load will be balanced, resulting in much higher throughput compared to single \co group. This demonstrates the effectiveness of \co sharding in enhancing \co scalability.}

\vspace{-2mm}
\subsection{Detailed Study of Commit Latency and \CC Length}\label{latency comparison}
In this experiment, we conducted a detailed analysis of transaction latency and \cc length among different approaches. The benchmark we used was YCSB+T workload A (50\% write and 50\% read), with each transaction accessing all three shards. The experiment used the default 3-replica deployment. The number of clients remained fixed at 150, distributed evenly among all three datacenters. The Zipf coefficient was set to 0.7, and the concurrency control protocol used was OCC.


We first define three kinds of time to analyze the latency clearly. 
$L_{i-j}$ denotes the RTT between datacenter $i$ and $j$. $L_{i-max}$ refers to the maximum time required for datacenter $i$ to receive replies from all datacenters, determined by the network latency with the most distant datacenter. $L_{i-majority}$ indicates the time required for datacenter $i$ to receive replies from a majority of datacenters. 
The comparison takes the transaction initiated at datacenter $a$ as an example.

\vspace{1mm}

\noindent \textbf{Latencies.}
Under the 3-replica deployment, for \ourwork and Carousel, the transaction needs to interact with all other datacenters to get the votes, making the commit latency determined by $L_{a-max}$; \occ involves communicating with participant datacenters $i$ and requires waiting for datacenter $i$ to complete replication. Thus, the overall commit latency is $\max \limits_{i\in p} (L_{a-i}+L_{i-majority})$, where $p$ is the set of participant datacenters.

Based on the results in Fig.~\ref{latency}, we have two observations. Firstly, both \ourwork and Carousel demonstrate significantly lower commit latencies, since both of them are optimized for latency than \occ. Secondly, the transaction latencies vary depending on the datacenter where the transaction originates. This disparity is due to the differences in $L_{i-max}$ and $L_{i-majority}$ among different datacenters. Taking Hangzhou as an example and referring to Table~\ref{network latency}, it can be observed that its $L_{i-max}$ is approximately 230 ms, while the $L_{i-max}$ of San Francisco is around 140 ms. Consequently, the commit latency for transactions initiated in San Francisco is lower.

\vspace{1mm}
\noindent \textbf{\Cc lengths.}
Table~\ref{latency analysis table} presents an overview of the \cc length for different approaches. We assume that the transaction initiated at datacenter $a$, and analyze the \cc length of shard leaders at datacenter $b$. 

Theoretically, the \cc lengths of Carousel and \occ correspond to their commit latencies. As shown in Table~\ref{latency analysis table}, their \cc lengths are 1 \cross RTT and 2 \cross RTTs, respectively. Consequently, their actual \cc lengths follow the same increasing order, as depicted in Fig.\ref{cc}(b) and (c). 

In contrast, \ourwork stands out with the shortest \cc length, requiring only 0.5 \cross RTT. However, the \cc length of \ourwork exhibits significant variation across different datacenters.
We set the timing of commit initiation as 0. For the shard leader in datacenter $b$, the \cc starts when it receives the \prepare message at time $\frac{L_{a-b}}{2}$, and concludes when receives all vote messages from other participant shards at the time $\max \limits_{i\in p} \frac{L_{a-i}+L_{i-b}}{2}$, where $p$ is the set of participant datacenters. Therefore, the \cc length is $\max \limits_{i\in p} \frac{L_{a-i} + L_{i-b} - L_{a-b}}{2}$. 

For example, taking Hangzhou as the initiated datacenter, referring to the network latencies in Table~\ref{network latency}, the \cc lengths for leaders in Frankfurt are only about 30 ms, which is much smaller than the \cc lengths in other datacenters, as depicted in Fig.~\ref{cc}(a). As a result, the \cc length in \ourwork exhibits variation depending on the locations. However, the overall \cc length of \ourwork is still significantly lower than other approaches.

\begin{table}[t]
\renewcommand\arraystretch{1.5}
	\centering
	\caption{Analysis of \cc length of shard leaders at $b$ (transaction initiated at $a$).}
	\label{latency analysis table}
 \vspace{-2mm}
	\begin{tabular}{ccc}
    \toprule[1.5pt]
	\docc	& Carousel & \occ \\
	\midrule
    $\max \limits_{i\in p} \frac{L_{a-i} + L_{i-b} - L_{a-b}}{2}$ & $L_{a-max}$ & $\max \limits_{i\in p} (L_{a-i}+L_{i-majority})$ \\
    \hline
    \toprule[1.5pt]
	\end{tabular}
\end{table}



\section{Conclusion}\label{conclusion}
This paper proposed the decentralized transaction commit protocol that offers several key insights. The primary objective is to minimize the impact of cross-region communication on system currency and commit latency. To achieve this, \ourwork leverages decentralized commit via \coss, and parallels processes of 2PC and replication. 
Compared to commit approaches based on 2PC, \ourwork demonstrates significant performance improvements in terms of throughput and latency. Furthermore, when compared to leaderless commit approaches that achieve fast commit, \ourwork realizes a similar reduction in commit latency while also delivering much higher improvements in throughput. Our experimental study demonstrated these promising characteristics.



\bibliographystyle{ACM-Reference-Format}
\bibliography{main}

\end{document}